\documentclass{aa}
\usepackage{graphics}
\begin{document}
\newcommand \phd \phantom{.}
\newcommand \phn \phantom{0}

\thesaurus{09.16.1, 10.01.1, 10.03.1}

\title{Observations of planetary nebulae in the Galactic Bulge 
\thanks{based on observations made at ESO, La Silla, Chile}}

\author{F. Cuisinier \inst{1}
\and W.J. Maciel \inst{2}
\and J. K\"oppen \inst{3}
\and A. Acker \inst{4}
\and B. Stenholm \inst{5}}

\institute{Observatorio do  Valongo, Depto. de Astronomia, UFRJ, 
         Ladeira do Pedro Antonio 43,
 20080~--~090~Rio~de~Janeiro~RJ, Brazil
\and
Instituto Astron\^omico e Geofisico da USP, Depto. de Astronomia, CP 9638, 
01065~--~970~S\~ao~Paulo~SP, Brazil
\and
Institut f\"ur theor. Physik u. Astrophysik der 
Universit\"at zu Kiel, 24~098 Kiel, Germany
\and
Observatoire de Strasbourg, 11 rue de l' Universit\'e, 67~000 Strasbourg,
 France
\and
Lund Observatory, Box 43, 221~00, Lund, Sweden}

\offprints{F. Cuisinier}

\date {Received June 8; Accepted September 30, 1999}

\maketitle

\begin{abstract}
High quality spectrophotometric observations of 30 Planetary Nebulae 
in the Galactic Bulge have been made. Accurate reddenings, plasma parameters,
and abundances of He, O, N, S, Ar, Cl are derived.

We find the abundances of O, S, Ar in 
the Planetary Nebulae in the Galactic Bulge 
to be comparable with the abundances of the Planetary Nebulae in the Disk,
 high abundances being maybe slightly more frequent in the Bulge.

The distribution of the N/O ratio does not present in the Galactic Bulge Planetary 
Nebulae the extension to 
high values that it presents  in the Disk Planetary Nebulae. We 
interpret this as a signature of the greater age of Bulge Planetary Nebulae.

We thus find the Bulge Planetary Nebulae to be an old population, slightly more 
metal-rich than the Disk Planetary Nebulae. 
The population of the Bulge Planetary Nebulae shows 
hence the same  characteristics than
the Bulge stellar population.

\end{abstract}

\section{Introduction}

 Color magnitude diagrams in the Bulge (Ortolani  1998) show  without
any doubt that  the  Bulge is an old population --- maybe the oldest
in the Galaxy. 

  Chemical abundances in the Bulge can be accessed through tracer populations
--- Red Giant stars, RR Lyrae and Planetary Nebulae. Since RR Lyrae stars are 
variable stars occurring only for a certain range of metallicities, their
metallicity distribution function is highly biased. On the other hand, 
red giant stars 
 show a distribution of metallicities as high as that for  the solar 
neighbourhood, or maybe higher (McWilliam \& Rich 1994). 

Planetary Nebulae (PN) are interesting objects for the study of the  Galactic Bulge 
because they concentrate the energy of 
their central stars in the emission lines of their spectra, 
 and can  therefore  be observed relatively easily
at this distance. Some of the elements that can be observed in 
Planetary Nebulae, such as O, S, Ar,  
reflect the composition of the interstellar 
medium when the progenitor star was born --- and can thus directly be compared
with  stellar abundances, mainly derived for Fe. Other elements,
 such as N and He are
synthetized in the PN progenitor stars and thus have abundances that are
contaminated by the products of this synthesis. These abundances should 
therefore be  indicative of the PN progenitor star {\em masses}, and
 therefore {\em ages}. Ratag et al. (1992, 1997) analysed the spectra 
of a sample of Galactic Bulge Planetary Nebulae. They found that the oxygen abundance 
distribution was  similar  or maybe  slightly deficient when compared with the 
solar neighbourhood one --- which is compatible with today's knowledge of the 
Galactic Bulge star abundances. They found as well that the    N/O ratio
was higher for Galactic Bulge Planetary Nebulae than for the Disk ones, which
requires (1) that the Galactic Bulge Planetary Nebulae progenitors were 
younger than the Disk ones, in contradiction with the stellar data, or (2)
that the PN progenitor star nucleosynthesis history was different in the 
Bulge and in the Disk.
  
  The Ratag et al. (1992, 1997) sample is however of very uneven quality: 
in many spectra, important diagnostic lines as [OIII] 436.2~nm or 
[NII] 575.5~nm are missing. This motivated us to obtain a sample of 
high quality PN spectra.

\section{Observations and reductions}

Belonging to the Galactic Bulge can be a tricky question, because of the contamination 
of foreground objects. For Planetary Nebulae, clear criteria have been identified
by Acker et al. (1991),
 that, according to them, allow to reduce the contamination 
of foreground objects down to less than 10\%. These criteria are: 
(i) diameter smaller than 10 arcsec, (ii) coordinates within 5 degrees of the
galactic center, and (iii) flux at 6cm less than 100 mJy. We have built a sample of Planetary
Nebulae  in the direction of the Bulge  responding as far as possible to the criteria 
identified  by Acker et al. (1991).\\

We obtained  spectra for a sample of 30 Planetary Nebulae 
responding to these criteria.
 The only PN of our sample that 
does not respond to these criteria is  PN~G~351.1~+~04.8 (M~1~-~19), 
but although it is not 
within 5 degrees of the galactic center, it is quite close, and it agrees with
 all the other criteria. Thus, we can assume that most of the PN of our 
sample actually belong to 
the Galactic Bulge.\\

The observations were made at the European Southern
Observatary  1.52m telescope
in La Silla, Chile, in two observing runs, in July  1995 and in July 1996.
 The instrument used was the B\&C  spectrograph 
with the grating \#23, and the CCD 2K chip FA\#24, giving a spectral coverage
of the range 380--750~nm, with a resolution of 0.6~nm. Such a resolution
allowed us to separate clearly the [SII] 671.6, 673.1~nm lines, 
and [OIII] 436.3~nm, 
${\rm H\beta}$.
The  sky apperture we used was of $3\times 4$arcsec.\\

Because of the high interstellar reddening towards
the Galactic Bulge, high quality spectra are difficult to obtain: weak lines 
that are  indispensable for the plasma diagnosis, like  [OIII] 436.3~nm or 
[NII] 575.5~nm are so reddenned that they cannot be observed in good conditions
(let us say, at S/N$\geq 10$) without saturating the most intense ones, as
H$\alpha$ or [OIII] 500.7~nm. 
 
Thus, for each PN, we obtained (i) one  spectrum with the weak lines, at least
[NII] 575.5~nm, observed in good conditions, which required exposures
times varying between 45  and 90 mn, (ii) one spectrum with the  
strongest lines unsaturated, typically with exposure  times of 5 minutes.

Each night 3--4 observations of one or several of the following 
spectrophotometric standard flux stars
 were made:
LTT 377, LTT 7379, Feige 110, CD -32 9927, and Kopff 27.
References of their fluxes distributions were taken from 
Baldwin \& Stone (1984),
Stone \& Baldwin (1983),
Colina \& Bohlin (1994), Hamuy et al. (1992, 1994), Massey et al. (1988), Oke (1990),
Stone (1977, 1996). 

The spectra have been reduced in wavelength and in flux with the MIDAS 
 package. The lines have been measured with the IRAF package.   
Some of the lines we observed, although separated, suffered  partial
blending: [SII] 671.6, 673.1~nm and [OII] 732,733~nm (as a matter of fact
[OII] 732,733~nm is a quadruplet: 731.8, 731.9, 733.0, 733.1~nm.
 But at our resolution, 
we were only able to separate the  two doublets). 
These lines were properly  deconvolved for partial blending 
with multiple gaussian fitting. [OIII] 436.3~nm and  ${\rm H\beta}$ were 
clearly separated, so  they did not require any deblending.

The line intensities are presented in Table \ref{tablelineint}.

\begin{table*}
\caption[]{Line intensities of the observed planetary nebulae, relatively to 
${\rm H\beta=100}$. Quality classes are  presented as well (see text).} 
\label{tablelineint}
\scriptsize
\begin{tabular}{llrrrrrrrr}
\noalign{\smallskip} \hline \noalign{\smallskip} 
$\lambda\lambda$ &  ident. &
$000.1+04.3$ & $000.2-01.9$ & 
$000.7+03.2$ & $001.2-03.0$ &
$001.4+05.3$ & $002.6-03.4$ &
$002.7-04.8$ & $005.0+04.4$ \\
${\rm (\AA)}$ & qual.: &
C & A &
C & A &
B & A & 
A & A \\
\noalign{\smallskip} \hline  \noalign{\smallskip}
4026 &  He I        &   -    &    -    &     -    &     -    &     -      &   -   &      -     &    1.3        \\
4072 & [S II]       &   -    &    -    &     -    &     -    &     -      &   -   &      -     &    2.1        \\
4102 &  H  I        &   -    &   10.8  &     -    &    12.6  &    13.6    &  13.6 &     17.7   &   11.1        \\
4190 &  O  II       &   -    &    -    &     -    &     -    &     -      &   -   &      -     &    -          \\
4267 &  C  II       &   -    &    -    &     -    &     -    &     -      &   -   &      -     &    0.5        \\
4340 &  H  I        &  18.2  &   27.1  &     -    &    27.7  &    28.7    &  29.2 &     33.9   &   24.8        \\
4363 & [O  III]     &   6.6  &    -    &     -    &     -    &     -      &   -   &      -     &    1.0        \\
4388 &  He I        &   -    &    -    &     -    &     -    &     -      &   -   &      -     &    0.5        \\
4472 &  He I        &   -    &    3.8  &     -    &     -    &     -      &   -   &      5.8   &    4.9        \\
4639 & O  II        &   -    &    -    &     -    &     -    &     -      &   -   &      -     &    -          \\
4686 & He II        &  10.5  &    -    &    29.6  &     -    &     -      &   -   &      9.9   &    -          \\
4711 & [Ar IV]+blend &   -   &     -   &     -    &     -    &     -      &   -   &      -     &    1.8        \\
4740 & [Ar IV]      &   4.4  &    -    &     3.6  &     -    &     -      &   -   &      1.0   &    0.6        \\
4861 & H I & 100\phd\phn & 100\phd\phn & 100\phd\phn & 100\phd\phn & 100\phd\phn & 100\phd\phn & 100\phd\phn & 100\phd\phn \\
4922 & He I         &    -   &     -   &      -   &     -   &      -      &   -   &      2.0    &    -         \\
4959 & [O  III] &  678\phd\phn & 79.5 & 281\phd\phn & - &  130\phd\phn &  - &  170\phd\phn & 254\phd\phn \\
5007 & [O  III] & 2171\phd\phn & 242\phd\phn & 882\phd\phn & - & 315\phd\phn &  - & 519\phd\phn & 808\phd\phn \\
5199 & [N  I]   &        -   &     0.6 &      -   &     1.5 &      -      &   1.6 &      3.7   &    0.7        \\
5271 & [FeII],[FeIII] &  -   &     -   &      -   &      -  &      -      &   -   &      -     &    -          \\
5399 &  0  IV         &  -   &     -   &      -   &      -  &      -      &   0.2 &      -     &    -          \\
5412 &  He II   &        3.3 &     -   &      5.5 &      -  &      -      &   -   &      1.3   &    -          \\
5518 & [Cl III] &       -    &    0.8  &     -    &     -   &      -      &   -   &      0.8   &    0.3        \\
5538 & [Cl III] &       -    &    0.8  &     -    &     -   &      -      &   -   &      0.8   &    1.5        \\
5603 &  ?       &       -    &    -    &     -    &     -   &      -      &   -   &      -     &    -          \\
5614 &  ?       &       -    &    -    &     -    &     -   &      -      &   -   &      -     &    -          \\
5667 &  N  II   &       -    &    -    &     -    &     -   &      -      &   -   &      0.7   &    -          \\
5676 &  N  II   &       -    &    -    &     -    &     -   &      -      &   -   &      -     &    -          \\
5696 &  C  III  &       -    &    -    &     -    &     2.4 &      -      &   3.2 &      -     &    -          \\
5754 & [N  II]  &       5.0  &    2.8  &     5.5  &     1.3 &      0.8    &   1.3 &      4.1   &    5.7        \\
5773 &  C  III  &       0.6  &    -    &     -    &     -   &      -      &   -   &      -     &    -          \\
5801 &  C  IV   &       -    &    -    &     -    &     -   &      -      &   -   &      -     &    -          \\
5812 &  C  IV   &       -    &    -    &     -    &     -   &      -      &   -   &      -     &    -          \\
5876 &  He I    &      52.8  &   37.9  &    47.4  &     3.1 &     33.8    &   3.2 &     33.7     &   60.7        \\
5932 &  He II   &       -    &    -    &     -    &     -   &      -      &   -   &      0.3   &    -          \\
6046 &  O  I?   &       -    &    -    &     -    &     -   &      -      &   0.3 &      -     &    -          \\
6102 & [K  IV]  &       -    &    -    &     -    &     -   &      -      &   -   &      -     &    -          \\
6300 & [O  I]   &      40.9  &    6.8  &    20.1  &     3.4 &      -      &   2.7 &      6.7   &   10.6        \\
6312 & [S  III] &      10.8  &    2.8  &     7.3  &     -   &      1.8    &   -   &      2.0   &    5.6        \\
6347 &  Si II   &        -   &     -   &      -   &     -   &      -      &   -   &      -     &    0.4        \\
6364 & [O  I]   &      15.7  &    2.4  &     7.7  &     1.1 &      -      &   1.0 &      2.6   &    3.9        \\
6437 & [Ar V]   &       2.2  &    -    &     -    &     -   &      -      &   -   &       -    &    -          \\ 
6462 &  ?       &        -   &    -    &     -    &     -   &      -      &   -   &      0.5   &    -          \\
6548 & [N II] & 76.0 & 147\phd\phn & 308\phd\phn & 230\phd\phn & 36.4\phd\phn & 191\phd\phn & 150\phd\phn & 159\phd\phn \\
6563 & H I & 1836\phd\phn & 1038\phd\phn & 1407\phd\phn & 954\phd\phn & 848\phd\phn & 700\phd\phn & 553\phd\phn & 1179\phd\phn \\
6583 & [N II] & 240\phd\phn & 464\phd\phn & 1008\phd\phn & 716\phd\phn & 115\phd\phn & 600\phd\phn & 461\phd\phn & 512\phd\phn\ \\
6678 & He I     &       27.9 &    16.4 &     26.3 &      -  &      13.2   &    -  &      12.4  &    25.9       \\
6716 & [S II]   &       18.5 &    39.1 &     62.0 &     27.2 &      5.9   &   16.1  &    30.0  &     8.0       \\
6731 & [S II]   &       35.9 &    56.1 &     94.5 &     53.4 &      7.3   &   32.2  &    37.1  &    17.3       \\
7006 & [Ar V]   &        8.0 &     -   &      4.0 &      0.5 &      -     &    0.4  &     -    &     -         \\
7065 &  He I      &     78.7 &    17.4 &     26.3 &      1.6 &     15.8   &    1.4  &     8.3  &    51.0       \\
7136 & [Ar III] & 178\phd\phn & 54.6 & 189\phd\phn & 1.3 & 46.3 & 0.7 &  40.8 & 132\phd\phn \\
7178 &  He II   &       -    &    -    &     -    &     -    &     -      &   -     &    -     &    -          \\
7281 &  He I    &       -    &    -    &     -    &     -    &     -      &   -     &    -     &    4.3        \\
7320 & [O  II] & 120\phd\phn &   17.4  &    20.1  &     2.5  &    13.8    &   1.9   &    5.4   &   20.1        \\
7330 & [O  II] &      95.4   &   14.4  &    16.1  &     1.8  &    11.3    &   1.4   &    4.9   &   17.0        \\
\noalign{\smallskip} \hline
\end{tabular}
\end{table*}

\begin{table*}
\addtocounter{table}{-1}
\caption[]{(continued)}
\scriptsize
\begin{tabular}{llrrrrrrrr}
\noalign{\smallskip} \hline \noalign{\smallskip}
$\lambda\lambda$ & ident. &
$351.1+04.8$ &  $355.1-02.9$ &
$355.2-02.5$ &  $355.4-02.4$ & 
$355.6-02.7$ &  $355.7-03.0$ &
$356.2-04.4$ &  $356.5-03.9$ \\
${\rm (\AA)}$ & qual.: &
A & A &
A & B &
A & A & 
A & A \\
\noalign{\smallskip} \hline \noalign{\smallskip}
4026 &  He I   &        -    &     -   &      -   &      -   &      -     &    1.2  &     -    &     -         \\
4072 &  [S II] &        1.5  &     -   &      1.7 &      -   &      -     &    2.9  &     2.4  &     -         \\
4102 &  H  I   &       16.8  &    11.0 &     12.4 &     12.2 &     11.1   &   12.7  &    18.6  &    12.3       \\
4190 &  O  II  &        -    &     -   &      -   &      -   &      0.8   &    -    &     -    &     -         \\
4267 &  C  II  &        0.3  &     -   &      -   &      -   &      -     &    0.9  &     -    &     -         \\
4340 &  H  I   &       34.4  &    24.9 &     28.3 &     26.1 &     25.8   &   27.9  &    35.5  &    29.2       \\
4363 & [O III] &        1.0  &     9.3 &      3.6 &      2.0 &      3.9   &    1.9  &     7.8  &     -         \\
4388 &  He I   &        0.5  &     -   &      0.4 &      -   &      -     &    0.6  &     0.6  &     -         \\
4472 &  He I   &        4.7  &     3.6 &      3.7 &      4.4 &      3.6   &    4.3  &     4.5  &     1.4       \\
4639 &  O  II  &         -   &     -   &      -   &      -   &      -     &    -    &     -    &     -         \\
4686 &  He II  &        1.4  &     3.9 &      -   &     13.3 &      -     &    2.2  &     4.3  &     -         \\
4711 & [Ar IV]+blend &  1.4  &     -   &      1.5 &      -   &      1.4   &    -    &     -    &     -         \\
4740 & [Ar IV] &        -    &     4.6 &      -   &      2.5 &      -     &    1.8  &     2.9  &     -         \\1
4861 & H I & 100\phd\phn & 100\phd\phn & 100\phd\phn & 100\phd\phn & 100\phd\phn & 100\phd\phn & 100\phd\phn & 100\phd\phn \\
4922 &  He I   &        -    &     -   &      -   &      -   &      -     &    1.6  &     1.3  &     -         \\
4959 & [O III] & 169\phd\phn & 653\phd\phn & 371\phd\phn & 305\phd\phn & 330\phd\phn & 309\phd\phn & 556\phd\phn & 7.0 \\
5007 & [O III] & 520\phd\phn & 2031\phd\phn & 1237\phd\phn & 947\phd\phn & 1021\phd\phn & 982\phd\phn & 1693\phd\phn & 22.8 \\
5199 & [N  I]  &        0.2  &     -   &      1.0 &      3.3 &      -     &    0.9  &     0.3  &     -         \\
5271 & [FeII],[FeIII] & 0.2  &     -   &      -   &      0.4 &      0.2   &      -  &     0.1  &     -         \\
5399 &  0  IV  &        -    &     -   &      -   &      -   &      -     &      - &      -    &     -         \\
5412 &  He II  &        0.1  &     0.6 &      -   &      2.2 &      -     &      -  &      0.5 &      -        \\
5518 & [Cl III] &       0.5  &     0.4 &      0.7 &      1.1 &      0.5   &    0.8  &     0.5  &     0.3       \\
5538 & [Cl III] &       0.9  &     1.1 &      1.3 &      2.0 &      1.0   &    1.2  &     0.9  &     0.5       \\
5603 &  ?       &       -    &     -   &      -   &      0.4 &      -     &    -    &     -    &     -         \\
5614 &  ?       &       -    &     -   &      -   &      -   &      -     &    -    &     -    &     -         \\
5667 &  N  II  &        -    &     -   &      -   &      0.3 &      -     &    -    &     -    &     -         \\
5676 &  N  II  &        -    &     -   &      -   &      -   &      -     &    0.5  &     -    &     -         \\
5696 &  C  III &        -    &     -   &      -   &      -   &      -     &    -    &     -    &     0.7       \\
5754 & [N  II] &        1.5  &     4.2 &      4.1 &      6.8 &       2.3  &     3.2 &      1.3 &     3.0       \\
5773 &  C  III &        -    &     -   &      -   &      -   &      -     &    -    &     -    &     -         \\
5801 &  C  IV  &        0.8  &     -   &      -   &      -   &      -     &    -    &     -    &     -         \\
5812 &  C  IV  &        0.6  &     -   &      -   &      -   &      -     &    -    &     -    &     -         \\
5876 &  He I   &       30.0  &    41.2 &     41.5 &     49.5 &     37.3   &   44.5  &    28.2  &    13.0       \\
5932 &  He II  &        -    &     -   &      -   &      -   &      -     &    -    &     -    &     -         \\
6046 &  O  I?  &        -    &     -   &      -   &      -   &      -     &    -    &     -    &     -         \\
6102 & [K IV]  &        -    &     -   &      -   &      -   &      -     &    -    &     0.2  &     -         \\
6300 & [O  I]  &        2.1  &    24.0 &     19.3 &     21.7 &     10.8   &   13.8  &     6.1  &     4.2       \\
6312 & [S  III] &       2.4  &     7.3 &      4.8 &      6.0 &      4.8   &    4.2  &     3.5  &     1.4       \\
6347 &  Si II  &        0.2  &     -   &      -   &      -   &      -     &    -    &     -    &     -         \\
6364 & [O  I]  &        0.7  &     8.6 &      6.4 &      8.5 &      3.7   &    4.7  &     2.2  &     1.3       \\
6437 & [Ar V]  &        -    &     -   &      -   &      -   &      -     &    -    &     -    &     -         \\
6462 &  ?      &        -    &     -   &      -   &      -   &      -     &    -    &     -    &     -         \\
6548 & [N  II] &       58.1  &    51.7 &     91.8 & 328\phd\phn & 29.2    &   91.5  &    20.5  &   133\phd\phn \\
6563 & H I & 665\phd\phn & 1063\phd\phn & 981\phd\phn & 1055\phd\phn & 960\phd\phn & 954\phd\phn & 594\phd\phn & 768\phd\phn \\
6583 & [N II] & 182\phd\phn & 163\phd\phn & 294\phd\phn & 1010\phd\phn & 90.7 & 291\phd\phn & 63.0 & 419\phd\phn \\
6678 &  He I  &        10.7  &    16.8 &     16.4 &     22.2 &     14.9   &   17.5  &     9.2  &     5.0       \\
6716 & [S II] &         5.5  &     7.5 &     11.6 &     43.9 &      4.7   &   22.1  &     3.6  &    16.3       \\
6731 & [S II] &        10.6  &    15.6 &     20.5 &     73.0 &      9.0   &   36.8  &     7.0  &    31.8       \\
7006 & [Ar V] &         -    &     -   &      -   &      0.6 &      -     &    -    &     -    &     -         \\
7065 &  He I  &        13.2  &     -   &     28.5 &     31.9 &     41.7   &   19.2  &    16.5  &     7.6       \\
7136 & [Ar III] &      36.4  &     -   &     57.1 & 146\phd\phn &  58.6   &   64.4  &    32.7  &    12.0       \\
7178 & He II  &        -     &    -    &     -    &     -    &     -      &   -     &    -     &    -          \\
7281 & He I   &        1.6   &    -    &     3.4  &     -    &     -      &   2.4   &    -     &    -          \\
7320 & [O II] &        6.8   &   48.1  &    27.9  &    18.9  &    46.1    &  10.8   &    7.5   &   29.6        \\
7330 & [O II] &        5.8   &   38.8  &    23.5  &    16.1  &    39.8    &   9.4   &    6.3   &   26.0        \\
\noalign{\smallskip} \hline 
\end{tabular}
\end{table*}

\begin{table*}
\addtocounter{table}{-1}
\caption[]{(continued)}
\scriptsize
\begin{tabular}{llrrrrrrrr}
\noalign{\smallskip} \hline \noalign{\smallskip}
$\lambda\lambda$ & ident. &
$356.5-02.3$ & $357.4-03.5$ &
$357.4-03.2$ & $358.2+03.5$ &
$358.2+03.6$ & $358.2+04.2$ &
$358.3+03.0$ & $358.7-05.2$ \\
${\rm (\AA)}$ & qual.: &
A & A &
B & A &
A & C &
B & A  \\
\noalign{\smallskip} \hline \noalign{\smallskip}
4026 &  He I  &         -    &    -    &     1.7  &     -    &     -      &   -     &    -     &    -          \\
4072 & [S II] &         -    &    -    &     3.6  &     1.2  &     -      &   -     &    -     &    -          \\
4102 &  H  I  &         -    &    -    &    14.0  &    10.5  &    12.0    &  11.7   &    -     &   16.5        \\
4190 &  O  II &         -    &    -    &     -    &     -    &     -      &   -     &    -     &    -          \\
4267 &  C  II &         -    &    -    &     -    &     -    &     -      &   -     &    -     &    -          \\
4340 &  H  I  &        26.0  &    -    &    28.0  &    24.5  &    27.2    &  26.2   &    -     &   31.1        \\
4363 & [O  III] &       -    &    -    &     2.8  &     5.6  &     7.8    &   -     &    -     &    9.7        \\
4388 &  He I  &         -    &    -    &     -    &     -    &     3.4    &   -     &    -     &    -          \\
4472 &  He I  &         -    &    3.8  &     4.1  &     3.0  &     -      &   4.9   &    -     &    3.6        \\
4639 &  O  II &         -    &    -    &     -    &     -    &     -      &   -     &    -     &    -          \\
4686 &  He II &         -    &    -    &    15.3  &     1.2  &    11.3    &   -     &    3.2   &    9.8        \\
4711 & [Ar IV]+blend &  -    &    -    &     5.5  &     -    &     -      &   -     &    -     &    -          \\
4740 & [Ar IV] &        -    &    -    &     2.7  &     2.3   &    4.8    &   -     &    5.1   &    5.0        \\
4861 & H I & 100\phd\phn & 100\phd\phn & 100\phd\phn & 100\phd\phn & 100\phd\phn & 100\phd\phn & 100\phd\phn & 100\phd\phn \\
4922 &  He I  &         -    &    1.3  &     1.7  &     -    &     1.4    &   -     &    -     &    1.2        \\
4959 & [O III] &  - & 164\phd\phn & 375\phd\phn & 467\phd\phn & 567\phd\phn & 217\phd\phn & 633\phd\phn & 589\phd\phn \\
5007 & [O III] &  - & 540\phd\phn & 1171\phd\phn & 1442\phd\phn & 1763\phd\phn & 689\phd\phn &  1980\phd\phn & 1812\phd\phn \\
5199 & [N  I] &         -    &    -    &     1.9  &     -    &     -      &   -     &    -     &    0.6        \\
5271 & [FeII],[FeIII] & -    &    -    &     -    &     -    &     -      &   -     &    -     &    -          \\
5399 &  0  IV  &        -    &    -    &     -    &     -    &     -      &   -     &    -     &    -          \\
5412 &  He II  &        -    &    -    &     2.2  &     -    &     2.0    &   -     &    -     &    1.4        \\
5518 & [Cl III] &       -    &    0.6  &     1.1  &     0.5  &     0.8    &   -     &    -     &    0.6        \\
5538 & [Cl III] &       -    &    1.1  &     1.8  &     0.8  &     1.3    &   -     &    -     &    1.4        \\
5603 & ?        &       -    &    -    &     -    &     -    &     -      &   -     &    -     &    -          \\
5614 & ?        &       -    &    -    &     -    &     1.9  &     -      &   -     &    -     &    -          \\
5667 & N  II    &       -    &    -    &     -    &     -    &     -      &   -     &    -     &    -          \\
5676 & N  II    &       -    &    -    &     -    &     -    &     -      &   -     &    -     &    -          \\
5696 &  C  III  &       2.0  &    -    &     -    &     -    &     -      &   -     &    -     &    -          \\
5754 & [N  II]  &       2.4  &    2.0  &     6.6  &     1.2  &     3.5    &   3.9   &   12.4   &    2.9        \\
5773 &  C  III  &       -    &    -    &     -    &     -    &     -      &   -     &    -     &    -          \\
5801 &  C  IV   &       -    &    -    &     -    &     -    &     -      &   -     &    -     &    -          \\
5812 &  C  IV   &       -    &    -    &     -    &     -    &     -      &   -     &    -     &    -          \\
5876 &  He I    &       3.4  &   34.5  &    41.3  &    45.6  &    36.9    &  52.1   &   63.2   &   26.5        \\
5932 &  He II   &       -    &    -    &     -    &     -    &     -      &   -     &    -     &    -          \\
6046 &  O  I?   &       -    &    -    &     -    &     -    &     -      &   -     &    -      &   -          \\
6102 & [K  IV]  &       -    &    -    &     -    &     -    &     0.7    &   -     &    -     &    0.4        \\
6300 & [O  I]   &       4.4  &    3.6  &    19.1  &     8.7  &    18.6    &   7.8   &   44.4   &   13.1        \\
6312 & [S  III]   &     -    &    3.0  &     6.9  &     5.1  &     6.1    &   3.6   &   17.8   &    4.2        \\
6347 &  Si II   &       -    &    -    &     -    &     -    &     -      &   -     &    -     &    -          \\
6364 & [O  I]   &       1.7  &    1.2  &     6.8  &     2.9  &        6.4 &   -     &   16.2   &    4.6        \\
6437 & [Ar V]   &       -    &    -    &     -    &     -    &     0.7    &   -     &    -     &    -          \\
6462 &  ?       &       -    &    -    &     -    &     -    &     -      &   -     &    -     &    -          \\
6548 & [N II]   & 258\phd\phn & 53.2 & 238\phd\phn & 24.5 &  56.5 & 128\phd\phn & 140\phd\phn & 38.2    \\
6563 &  H I & 1221\phd\phn & 828\phd\phn & 966\phd\phn & 1440\phd\phn & 1033\phd\phn & 1204\phd\phn & 1786\phd\phn & 611\phd\phn \\
6583 & [N II] & 817\phd\phn & 168\phd\phn & 875\phd\phn & 76.3\phd\phn & 172\phd\phn & 407\phd\phn & 440\phd\phn & 118\phd\phn \\
6678 &  He I  &         -    &   12.8  &    16.4  &    22.2  &    15.4    &  23.6   &   29.5   &    9.0        \\
6716 & [S  II] &       42.5  &    6.1  &    29.3  &     5.8  &    11.2    &  13.7   &   13.7   &    6.9        \\
6731 & [S  II]  &      82.2  &   11.4  &    52.8  &    11.5  &    22.0    &  26.2   &   29.5   &   13.2        \\
7006 & [Ar V]   &       0.8  &    -    &     -    &     -    &     2.4    &   -     &    1.3   &    1.1        \\
7065 &  He I    &       0.7  &   26.7  &    22.6  &    37.8  &    31.8    &  27.7   &   92.7   &   18.5        \\
7136 & [Ar III] &       3.0  &   45.3  & 102\phd\phn & 14.9 &  76.7   & 108\phd\phn & 221\phd\phn & 40.9   \\
7178 &  He II   &       -    &    -    &     0.7  &     -    &     -      &   -     &    -     &    -          \\
7281 &  He I    &       -    &    -    &     2.6  &     4.6  &     -      &   -     &    -     &    -          \\
7320 & [O  II]  &      13.6  &   40.3  &    13.6  &    20.8  &    36.2    &  25.1   &  100     &   28.9        \\
7330 & [O  II]  &      11.8  &   32.0  &    10.7  &    16.2  &    30.3    &  19.7   &   83.4   &   22.0        \\
\noalign{\smallskip} \hline
\end{tabular}
\end{table*}

\begin{table*}
\addtocounter{table}{-1}
\caption[]{(continued)}
\scriptsize
\begin{tabular}{llrrrrrr}
\noalign{\smallskip} \hline \noalign{\smallskip}
$\lambda\lambda$ & ident. &
$358.9+03.2$ & $358.9+03.4$ &
$359.1-02.3$ & $359.3-03.1$ &
$359.7-02.6$ & $359.7-01.8$ \\
${\rm (\AA)}$ & qual.: &
B & B &
B & B &
B & C \\
\noalign{\smallskip} \hline \noalign{\smallskip}
4026 &  He I    &        -   &    -    &     -    &     -    &     -      &   -     \\
4072 &  [S II]  &        -   &    -    &     -    &     -    &     -      &   -     \\
4102 &  H  I    &        -   &    -    &    12.6  &    12.0  &     -      &   -     \\    
4190 &  O  II   &        -   &    -    &     -    &     -    &     -      &   -     \\
4267 &  C  II   &        -   &    -    &     -    &     -    &     -      &   -     \\
4340 &  H  I    &        -   &    -    &    29.7  &    27.8  &     -      &  20.8   \\  
4363 & [O III]  &        -   &    -    &     0.9  &     -    &     -      &  11.2   \\ 
4388 &  He I    &        -   &    -    &     -    &     -    &     -      &   -     \\
4472 &  He I    &        3.6 &    -    &     4.1  &     2.8  &     -      &   -     \\
4639 &  O  II   &        0.3 &    -    &     -    &     -    &     -      &   -     \\
4686 &  He II        &   1.8 &    -    &     -    &     -    &     -      &  22.5   \\
4711 & [Ar IV]+blend &   -   &    -    &     -    &     -    &     -      &   -     \\
4740 & [Ar IV]  &        -   &    -    &     -    &     -    &     -      &   -     \\
4861 & H I & 100\phd\phn & 100\phd\phn & 100\phd\phn & 100\phd\phn & 100\phd\phn & 100\phd\phn \\
4922 &  He I    &        -   &    -    &     -    &     1.4  &     -      &   -     \\
4959 & [O  III] & 365\phd\phn & 148\phd\phn & 230\phd\phn & 26.1 & 353\phd\phn & 500  \\
5007 & [O  III] & 1164\phd\phn & 480\phd\phn & 713\phd\phn & 85.2 & 1131\phd\phn & 1584\phd\phn \\
5199 & [N  I]   &        1.2 &    -    &     -    &     -    &     4.3    &   -     \\
5271 & [FeII],[FeIII] &  -   &    -    &     -    &     -    &     -      &   -     \\
5399 &  0  IV   &        -   &    -    &     -    &     -    &     -      &   -     \\
5412 &  He II   &        -   &    -    &     -    &     -    &     -      &   -     \\ 
5518 & [Cl III] &        0.9 &    -    &     -    &     0.3  &     -      &   -     \\
5538 & [Cl III] &        2.1 &    -    &     -    &     0.7  &     -      &   -     \\
5603 &  ?       &        -   &    -    &     -    &     -    &     -      &   -     \\
5614 &  ?       &        -   &    -    &     -    &     -    &     -      &   -     \\
5667 &  N II    &        -   &    -    &     -    &     -    &     -      &   -     \\
5676 &  N II    &        -   &    -    &     -    &     0.5  &     -      &   -     \\
5696 &  C III   &        -   &    -    &     -    &     1.0  &     -      &   -     \\
5754 & [N  II]  &        7.2 &   14.0  &     1.6  &     2.6  &     5.7    &   -     \\
5773 &  C  III  &        -   &    -    &     -    &     -    &     -      &   -     \\
5801 &  C  IV   &        -   &    -    &     -    &     -    &     -      &   -     \\
5812 &  C  IV   &        -   &    -    &     -    &     -    &     -      &   -     \\
5876 &  He I    &       64.2 &   69.5  &    35.9  &    34.4  &    66.2    &  34.8   \\
5932 &  He II   &        -   &    -    &     -    &     -    &     -      &   -     \\
6046 &  O  I?   &        -   &    -    &     -    &     -    &     -      &   -     \\
6102 & [K  IV]  &        -   &    -    &     -    &     -    &     -      &   -     \\ 
6300 & [O  I]   &       20.0 &   11.6  &     8.2  &     1.7  &    15.1    &   7.3   \\
6312 & [S  III] &        8.4 &    6.7  &     3.0  &     2.6  &    10.5    &   7.3   \\
6347 &  Si II   &        0.9 &    -    &     -    &     -    &     -      &   -     \\
6364 & [O  I]   &        7.5 &    3.7  &     2.3  &     -    &     6.2    &   -     \\
6437 & [Ar V]   &        -   &    -    &     -    &     -    &     -      &   -     \\
6462 &  ?       &        -   &    -    &     -    &     -    &     -      &   -     \\
6548 & [N  II]  & 294\phd\phn & 433\phd\phn & 65.0 & 176\phd\phn & 96.8   &   27.0  \\
6563 & H I & 1597\phd\phn & 1680\phd\phn & 1065\phd\phn & 1143\phd\phn & 1840\phd\phn & 1704 \\
6583 & [N  II]  & 941\phd\phn & 1321\phd\phn & 204\phd\phn & 567\phd\phn & 305\phd\phn & 82.0 \\
6678 & He I     &       31.0 &   32.9  &    15.8  &    14.5  &    31.5    &  18.5   \\
6716 & [S II]   &       41.2 &   17.7  &    21.3  &    16.9  &     7.8    &  17.4   \\
6731 & [S II]   &       73.7 &   37.8  &    27.0  &    28.7  &    17.1    &  21.3   \\
7006 & [Ar V]   &         -  &     -   &      -   &      -   &      -     &   3.4   \\
7065 & He I     &       60.6 &   77.4  &    16.2  &    14.0  &    86.5    &  25.3   \\
7136 & [Ar III] & 227\phd\phn & 232\phd\phn & 45.3 &   36.1  & 148\phd\phn & 71.5   \\
7178 &  He II   &        -   &    -    &     -    &     -    &     -      &   4.5   \\
7281 &  He I    &        -   &    -    &     -    &     2.5  &     -      &   -     \\
7320 & [O II]   &       34.3 &   71.3  &    19.0  &    17.6  &    68.0    &  18.0   \\
7330 & [O II]   &       28.1 &   56.1  &    15.8  &    14.1  &    55.0    &  14.6   \\
\noalign{\smallskip} \hline
\end{tabular}
\end{table*}

\section{Plasma diagnosis and abundance determinations}

\subsection{Presentation of our data}

The interstellar reddenings and the plasma parameters 
(electronic temperatures, electronic densities) were derived with the plasma
diagnosis code HOPPLA (K\"oppen et al. 1991).  The interstellar reddenings
were derived using the Balmer line ratios.  For all PN of our sample, 
${\rm H\alpha}$ and ${\rm H\beta}$ could be measured in good conditions,
 and in most cases, ${\rm H\gamma}$ and ${\rm H\delta}$  as well.
The dereddened Balmer lines fitted
always within 15\% of the case B predicted values.

 Electronic temperatures
have been derived for the [NII] zone and, when possible, for the [OIII] zone. 
When the [OIII]~436.3~nm was too weak to be measurable in good conditions,
the [OIII] zone electronic temperature was assumed to be the same 
as in the [NII] zone. 

The electronic densities were derived using the [SII] 671.6, 673.1~nm 
line ratio.\\

The adopted plasma parameters,  interstellar reddenings and  excitation 
classes
of the PN of this sample are presented in Table \ref{tableplasma}.\\

Elemental abundances were derived from the ionic abundances using ionisation
correction factors (ICF). The ICF we used are described in K\"oppen et al.
1991. The abundances we derived for the PN of our sample are presented in
Table \ref{tableabund}.

\begin{table*}
   \caption{Reddenings (c) and plasma parameters (temperatures, densities)
 of the observed spectra. These data are given after the PN G names, the commen names, and our  evaluation of the quality of the spectra. Excitation classes (EC) are as well given. All the parameters are followed
by our evaluations of the uncertainties by Monte--Carlo simulations ($1\sigma$). The temperatures are given 
in units of $10^4$K, for the [OIII] and [NII] zones. The densities, derived from the [SII] lines ratio, 
 are given in ${\rm 10^3 cm^{-3}}$.}
   \label{tableplasma}
   \footnotesize
   \begin{tabular}{llcrrrrrrrrrrrr}
   \noalign{\smallskip} \hline \noalign{\smallskip}
  PN G & usual name & qual.  
                    & c \ \ \ & ${\rm \sigma_c}$  \ \ \ & T(OIII) & ${\rm \sigma_{T(OIII)}}$  & 
                       T(NII) & ${\rm \sigma_{T(NII)}}$  &  n(SII) & ${\rm \sigma_{n(SII)}}$     & EC\\
 \noalign{\smallskip} \hline
   \noalign{\smallskip}
$000.1+04.3 $  & H 1-16    &  C & 2.50~~~& $\pm0.02$ &  1.00 & $\pm0.09$ &  1.41 & $\pm0.30$  &   7.30~~~& $\pm6.0$ &     6 \\
$000.2-01.9 $  & M 2-19    &  A & 1.77~~~& $\pm0.01$ &     - &       -   &  0.80 & $\pm0.05$  &   1.86~~~& $\pm0.1$ &     3 \\
$000.7+03.2 $  & He 2-250  &  C & 2.11~~~& $\pm0.03$ &     - &       -   &  0.80 & $\pm0.09$  &   2.29~~~& $\pm0.4$ &     7 \\
$001.2-03.0 $  & H 1-47    &  A & 1.56~~~& $\pm0.01$ &     - &       -   &  0.56 & $\pm0.04$  &   7.04~~~& $\pm1.1$ & $<$ 2 \\
$001.4+05.3 $  & H 1-15    &  B & 1.45~~~& $\pm0.04$ &     - &       -   &  0.83 & $\pm0.27$  &   1.16~~~& $\pm1.8$ &     4 \\
$002.6-03.4 $  & M 1-37    &  A & 1.15~~~& $\pm0.01$ &     - &       -   &  0.56 & $\pm0.04$  &   8.13~~~& $\pm2.7$ & $<$ 2 \\
$002.7-04.8 $  & M 1-42    &  A & 0.88~~~& $\pm0.01$ &     - &       -   &  0.86 & $\pm0.04$  &   1.17~~~& $\pm0.1$ &     6 \\
$005.0+04.4 $  & H 1-27    &  A & 1.89~~~& $\pm0.01$ &  0.74 & $\pm0.09$ &  0.89 & $\pm0.04$  &  19.27~~~& $\pm4.7$ &     5 \\

   \noalign{\smallskip}
$351.1+04.8 $  & M 1-19    &  A & 1.11~~~& $\pm0.01$ &  0.78 & $\pm0.09$ &  0.82 & $\pm0.09$  &   6.76~~~& $\pm5.8$ &     4 \\
$355.1-02.9 $  & H 1-31    &  A & 1.78~~~& $\pm0.01$ &  1.05 & $\pm0.02$ &  1.38 & $\pm0.11$  &  12.47~~~& $\pm5.2$ &     6 \\
$355.2-02.5 $  & H 1-29    &  A & 1.65~~~& $\pm0.01$ &  0.92 & $\pm0.04$ &  1.07 & $\pm0.05$  &   4.41~~~& $\pm1.2$ &     5 \\
$355.4-02.4 $  & M 3-14    &  B & 1.74~~~& $\pm0.01$ &  0.85 & $\pm0.10$ &  0.82 & $\pm0.04$  &   3.30~~~& $\pm0.4$ &     6 \\
$355.6-02.7 $  & H 1-32    &  A & 1.64~~~& $\pm0.01$ &  0.99 & $\pm0.04$ &  1.41 & $\pm0.19$  &   7.45~~~& $\pm6.0$ &     5 \\
$355.7-03.0 $  & H 1-33    &  A & 1.61~~~& $\pm0.01$ &  0.82 & $\pm0.04$ &  0.98 & $\pm0.06$  &   3.38~~~& $\pm0.4$ &     5 \\
$356.2-04.4 $  & Cn 2-1    &  A & 0.98~~~& $\pm0.01$ &  0.98 & $\pm0.04$ &  1.17 & $\pm0.31$  &   6.85~~~& $\pm8.1$ &     6 \\
$356.5-03.9 $  & H 1-39    &  A & 1.31~~~& $\pm0.01$ &   -   &       -   &  0.79 & $\pm0.04$  &   7.50~~~& $\pm2.3$ &     2 \\
   \noalign{\smallskip}
$356.5-02.3 $  & M 1-27    &  A & 1.90~~~& $\pm0.03$ &   -   &       -   &  0.64 & $\pm0.11$  &   6.53~~~& $\pm3.2$ & $<$ 2 \\
$357.4-03.5 $  & M 2-18    &  B & 1.41~~~& $\pm0.01$ &   -   &       -   &  0.95 & $\pm0.09$  &   7.47~~~& $\pm4.2$ &     6 \\
$357.4-03.2 $  & M 2-16    &  B & 1.62~~~& $\pm0.01$ &  0.87 & $\pm0.08$ &  0.85 & $\pm0.04$  &   4.81~~~& $\pm1.1$ &     6 \\
$358.2+03.5 $  & H 2-10    &  A & 2.18~~~& $\pm0.01$ &  1.04 & $\pm0.05$ &  1.16 & $\pm0.24$  &   8.39~~~& $\pm6.0$ &     5 \\
$358.2+03.6 $  & M 3-10    &  A & 1.73~~~& $\pm0.01$ &  1.04 & $\pm0.02$ &  1.24 & $\pm0.10$  &   8.20~~~& $\pm3.6$ &     6 \\
$358.2+04.2 $  & M 3-8     &  C & 1.92~~~& $\pm0.03$ &   -   &       -   &  0.92 & $\pm0.18$  &   6.51~~~& $\pm6.7$ &     4 \\
$358.3+03.0 $  & H 1-17    &  B & 2.48~~~& $\pm0.01$ &   -   &       -   &  1.52 & $\pm0.11$  &  18.18~~~& $\pm4.7$ &     6 \\
$358.7-05.2 $  & H 1-50    &  A & 1.04~~~& $\pm0.01$ &  1.04 & $\pm0.02$ &  1.28 & $\pm0.13$  &   7.40~~~& $\pm5.0$ &     6 \\
   \noalign{\smallskip}
$358.9+03.2 $  & H 1-20    &  B & 2.29~~~& $\pm0.01$ &    -  &       -   &  0.89 & $\pm0.04$  &   4.59~~~& $\pm0.7$ &     5 \\
$358.9+03.4 $  & H 1-19    &  B & 2.36~~~& $\pm0.01$ &    -  &       -   &  0.93 & $\pm0.04$  &  15.85~~~& $\pm4.9$ &     4 \\
$359.1-02.3 $  & M 3-16    &  B & 1.73~~~& $\pm0.02$ &  0.75 & $\pm0.16$ &  0.89 & $\pm0.16$  &   1.22~~~& $\pm0.3$ &     4 \\
$359.3-03.1 $  & M 3-17    &  B & 1.83~~~& $\pm0.02$ &   -   &       -   &  0.74 & $\pm0.08$  &   3.46~~~& $\pm1.2$ &     3 \\
$359.7-02.6 $  & H 1-40    &  B & 2.50~~~& $\pm0.02$ &   -   &       -   &  1.18 & $\pm0.13$  &  20.00~H &    -     &     5 \\
$359.7-01.8 $  & M 3-45    &  C & 2.42~~~& $\pm0.03$ &  1.35 & $\pm0.93$ &   -   &     -      &   1.25~~~& $\pm0.9$ &     6 \\
   \noalign{\smallskip} \hline
   \end{tabular}
\end{table*}

\begin{table*}
   \caption[]{Abundances of the observed PN. The abundances are given on a logarithmic scale, where 12 is 
             the abundance of hydrogen. All the abundances are  followed by our evaluations of the uncertainties
             from Monte--Carlo simulations ($1\sigma$). Abundances judged of low quality, 
because of a lack of data, are followed by a semi-colon (:).  As well, the excitation classes EC are mentioned.}
   \label{tableabund}
   \footnotesize
   \begin{tabular}{lrrrrrrrrrrrrr}
   \noalign{\smallskip} \hline \noalign{\smallskip}
    PN G        & EC& He \ \ & ${\rm \sigma_{He}}$  \ \ & N \ \ & ${\rm \sigma_N}$ \ \ & O \ \ & ${\rm \sigma_O}$  \ \ &
      S \ \ & ${\rm \sigma_S}$ \ \ &  Ar \ \ & ${\rm \sigma_{Ar}}$ \ \ & Cl \ \ & ${\rm \sigma_{Cl}}$ \ \ \\
   \noalign{\smallskip} \hline
   \noalign{\smallskip}
$000.1+04.3 $&    6 & 10.97~~&  $\pm0.02$& 8.09~~& $\pm0.2$& 8.87~~& $\pm0.1$ & 6.69~~& $\pm0.2$&  6.44~~& $\pm0.1$ &   
  - ~~&       - \\
$000.2-01.9 $&    3 & 10.98~:&  $\pm0.01$& 7.89~~& $\pm0.1$& 8.61~~& $\pm0.2$ & 6.99~~& $\pm0.2$&  6.32~:& $\pm0.1$ & 
 5.03~&  $\pm0.3$ \\
$000.7+03.2 $&    7 & 11.09~~&  $\pm0.02$& 8.61~~& $\pm0.1$& 9.00~~& $\pm0.3$ & 7.30~~& $\pm0.3$&  6.85~~& $\pm0.2$ &
   - ~~&      - \\
$001.2-03.0 $& $<$ 2 &  9.89~:&      -   & 8.52~~& $\pm0.1$& 8.85~~& $\pm0.3$ & 6.92~:& $\pm0.1$&  5.53~:& $\pm0.3$ &
  -  ~~&      -  \\
$001.4+05.3 $&     4 & 11.00~:& $\pm0.03$& 7.43~~& $\pm0.3$& 8.69~~& $\pm1.0$ & 6.75~~& $\pm0.9$&  6.27~:& $\pm0.5$ & 
  -  ~~&      -  \\
$002.6-03.4 $& $<$ 2 & 10.01~:&      -   & 8.56~~& $\pm0.1$& 8.83~~& $\pm0.4$ & 6.86~:& $\pm0.2$&  5.47~:& $\pm0.3$ &
  -  ~~&      -  \\
$002.7-04.8 $&     6 & 11.18~~& $\pm0.01$& 8.40~~& $\pm0.1$& 8.65~~& $\pm0.1$ & 6.98~~& $\pm0.1$&  6.55~~& $\pm0.1$ &
 5.17~~& $\pm0.2$ \\
$005.0+04.4 $&     5 & 11.13~:& $\pm0.01$& 8.72~~& $\pm0.2$& 8.90~~& $\pm0.2$ & 7.11~~& $\pm0.1$&  6.70~~& $\pm0.1$ & 
 5.37~~& $\pm0.2$ \\
   \noalign{\smallskip}
$351.1+04.8 $&     4 & 11.03~~& $\pm0.01$& 8.19~~& $\pm0.2$& 8.73~~& $\pm0.2$ & 7.03~~& $\pm0.2$&  6.42~~& $\pm0.2$ &
 5.22~~& $\pm0.3$ \\
$355.1-02.9 $&     6 & 11.00~~& $\pm0.01$& 8.21~~& $\pm0.1$& 8.77~~& $\pm0.1$ & 6.77~~& $\pm0.1$&  5.86~:& $\pm0.1$ &
 4.77~:&   -       \\
$355.2-02.5 $&     5 & 11.03~:& $\pm0.01$& 8.21~~& $\pm0.1$& 8.75~~& $\pm0.1$ & 6.77~~& $\pm0.1$&  6.19~~& $\pm0.1$ & 
 5.02~:&   -       \\
$355.4-02.4 $&     6 & 11.13~~& $\pm0.01$& 8.61~~& $\pm0.2$& 8.90~~& $\pm0.2$ & 7.26~~& $\pm0.1$&  6.76~~& $\pm0.2$ & 
 5.39~~& $\pm0.3$  \\
$355.6-02.7 $&     5 & 10.98~:& $\pm0.01$& 7.68~~& $\pm0.1$& 8.55~~& $\pm0.1$ & 6.46~~& $\pm0.1$&  6.11~~& $\pm0.1$ &
 4.81~~& $\pm0.2$  \\
$355.7-03.0 $&     5 & 11.09~~& $\pm0.02$& 8.51~~& $\pm0.1$& 8.82~~& $\pm0.1$ & 6.99~~& $\pm0.1$&  6.53~~& $\pm0.1$ &
 5.16~~& $\pm0.2$  \\
$356.2-04.4 $&     6 & 11.04~~& $\pm0.01$& 8.28~~& $\pm0.2$& 8.80~~& $\pm0.1$ & 6.99~~& $\pm0.1$&  6.30~~& $\pm0.1$ & 
 4.93~~& $\pm0.2$  \\
$356.5-03.9 $&     2 & 10.60~~&     -    & 7.82~~& $\pm0.1$& 8.67~~& $\pm0.2$ & 6.90~~& $\pm0.2$&  5.94~:& $\pm0.1$ &
 4.95~~ & $\pm0.3$ \\
   \noalign{\smallskip}
$356.5-02.3 $& $<$ 2 &  9.88~:&     -    & 8.22~~& $\pm0.1$& 8.86~~& $\pm0.2$ & 6.74~:& $\pm0.1$&  5.44~:& $\pm0.3$ &
  -  ~~ &  -       \\
$357.4-03.5 $&     6 & 11.00~~& $\pm0.01$& 7.48~~& $\pm0.1$& 8.54~~& $\pm0.3$ & 6.68~~& $\pm0.2$&  6.12~~& $\pm0.1$ &
 4.94~~ & $\pm0.4$ \\
$357.4-03.2 $&     6 & 11.09~~& $\pm0.04$& 8.81~~& $\pm0.2$& 8.91~~& $\pm0.1$ & 7.30~~& $\pm0.1$&  6.64~~& $\pm0.1$ &
 5.37~~ & $\pm0.3$ \\
$358.2+03.5 $&     5 & 10.95~~& $\pm0.01$& 7.87~~& $\pm0.2$& 8.62~~& $\pm0.1$ & 6.74~~& $\pm0.2$&  5.78~~& $\pm0.1$ & 
 4.61~: &   -      \\
$358.2+03.6 $&     6 & 11.00~~& $\pm0.01$& 8.17~~& $\pm0.1$& 8.76~~& $\pm0.1$ & 6.74~~& $\pm0.1$&  6.37~~& $\pm0.1$ & 
 4.97~: &   -       \\
$358.2+04.2 $&     4 & 11.08~:& $\pm0.04$& 8.08~~& $\pm0.2$& 8.55~~& $\pm0.5$ & 6.75~~& $\pm0.5$&  6.36~~& $\pm0.3$ &
  -  ~~ &   -       \\
$358.3+03.0 $&     6 & 10.92~~& $\pm0.03$& 8.08~~& $\pm0.1$& 8.32~~& $\pm0.1$ & 6.62~~& $\pm0.1$&  6.09~~& $\pm0.1$ & 
  -  ~~ &   -       \\
$358.7-05.2 $&     6 & 11.01~~& $\pm0.01$& 8.09~~& $\pm0.1$& 8.80~~& $\pm0.1$ & 6.70~~& $\pm0.1$&  6.38~~& $\pm0.1$ &
 4.95~: &   -       \\
   \noalign{\smallskip}
$358.9+03.2 $&     5 & 11.06~~& $\pm0.01$& 8.47~~& $\pm0.1$& 8.80~~& $\pm0.1$ & 7.09~~& $\pm0.1$&  6.56~:& $\pm0.1$ &
 5.18~~ & $\pm0.3$  \\
$358.9+03.4 $&     4 & 11.07~:& $\pm0.01$& 8.24~~& $\pm0.1$& 8.47~~& $\pm0.1$ & 6.86~~& $\pm0.1$&  6.52~:& $\pm0.1$ &
  -  ~~ &   -       \\
$359.1-02.3 $&     4 & 10.97~:& $\pm0.02$& 7.88~~& $\pm0.3$& 8.90~~& $\pm0.4$ & 6.84~~& $\pm0.4$&  6.27~:& $\pm0.2$ &
  -  ~~ &   -      \\
$359.3-03.1 $&     3 & 10.90~:& $\pm0.03$& 7.94~~& $\pm0.2$& 8.65~~& $\pm0.5$ & 7.09~~& $\pm0.4$&  6.18~:& $\pm0.2$ &
 5.09~~ & $\pm0.4$  \\
$359.7-02.6 $&     5 & 10.99~:& $\pm0.02$& 7.90~:& $\pm0.1$& 8.36~:& $\pm0.2$ & 6.63~:& $\pm0.2$&  5.98~:& $\pm0.1$ &
  -  ~~ &   -       \\
$359.7-01.8 $&     6 & 10.93~~& $\pm0.02$& 7.71~~& $\pm0.1$& 8.45~~& $\pm0.1$ & 6.53~~& $\pm0.2$&  5.62~:& $\pm0.1$ &
  -  ~~ &  -       \\
   \noalign{\smallskip} \hline
   \end{tabular}
\end{table*}

\subsection{Internal  uncertainties}

Uncertainties on the measured spectra originate from various sources, random and systematic.
The random uncertainties can be divided between: (i) the uncertainty on the 
determination of the atmospheric transmission, (ii) the CCD read--out noise,
 and (iii) the 
photon shooting noise. Cuisinier et al. (1992) showed that the total derived
response of the instrument, with the atmospheric transmission, was constant 
within 5\% over 100nm.
 The exposure times were chosen appropriately, so that the CCD 
read-out noise can be neglected. Thus, the main random error source should be the 
photon shooting noise, which dominates the line measurement uncertainties, through the uncertainty on the continuum level setting.\\

In order to evaluate random uncertainties on the plasma parameters and 
abundances we derived, 
we performed Monte-Carlo simulations, adding gaussian noise to the observed 
spectra.
Random values have been added and subtracted to each of the lines of each of 
the observed observed spectra. For each spectrum, 200 shots have been made. 
  A visual inspection of our spectra showed  that the noise level 
was fairly constant over the wavelength range, showing 
only a notable increase at the 
ends of the range, due to the loss of efficiency of the spectrograph at high
grating angles.\\

The level of the gaussian noise we added to the spectra was   based  on our visual evaluation of
the uncertainty on the line measurements. We adopted standard deviations for the random values we added to the spectra reproducing 
our evaluation of the noise level.
 We divided the spectra in three
classes: A, B, and C. C having an uncertainty level of 2\% in ${\rm H\beta}$ 
units,
B 1\%, and A less than 0.5\%.\\ 

Table \ref{tableplasma} presents the standard deviations 
on each of the plasma parameters for each spectrum that we derived. 
For the density, in cases where the evaluated 
density reaches the saturation limits of the sulfur lines 
(taken as 100${\rm cm^{-3}}$ for the low density limit and 
20,000${\rm cm^{-3}}$ for the high density limit), these values of the
  standard deviations should just be seen as indications. No values 
have been generated  over the high-density limit (or below the low density 
one), resulting sometimes for the concerned spectra  
in highly non--gaussian distributions of the random generated densities. 

Table \ref{tableabund} presents the standard deviations on each of the 
abundances for each spectrum that we derived. 

\begin{figure*}
\resizebox{\hsize}{!}{\includegraphics{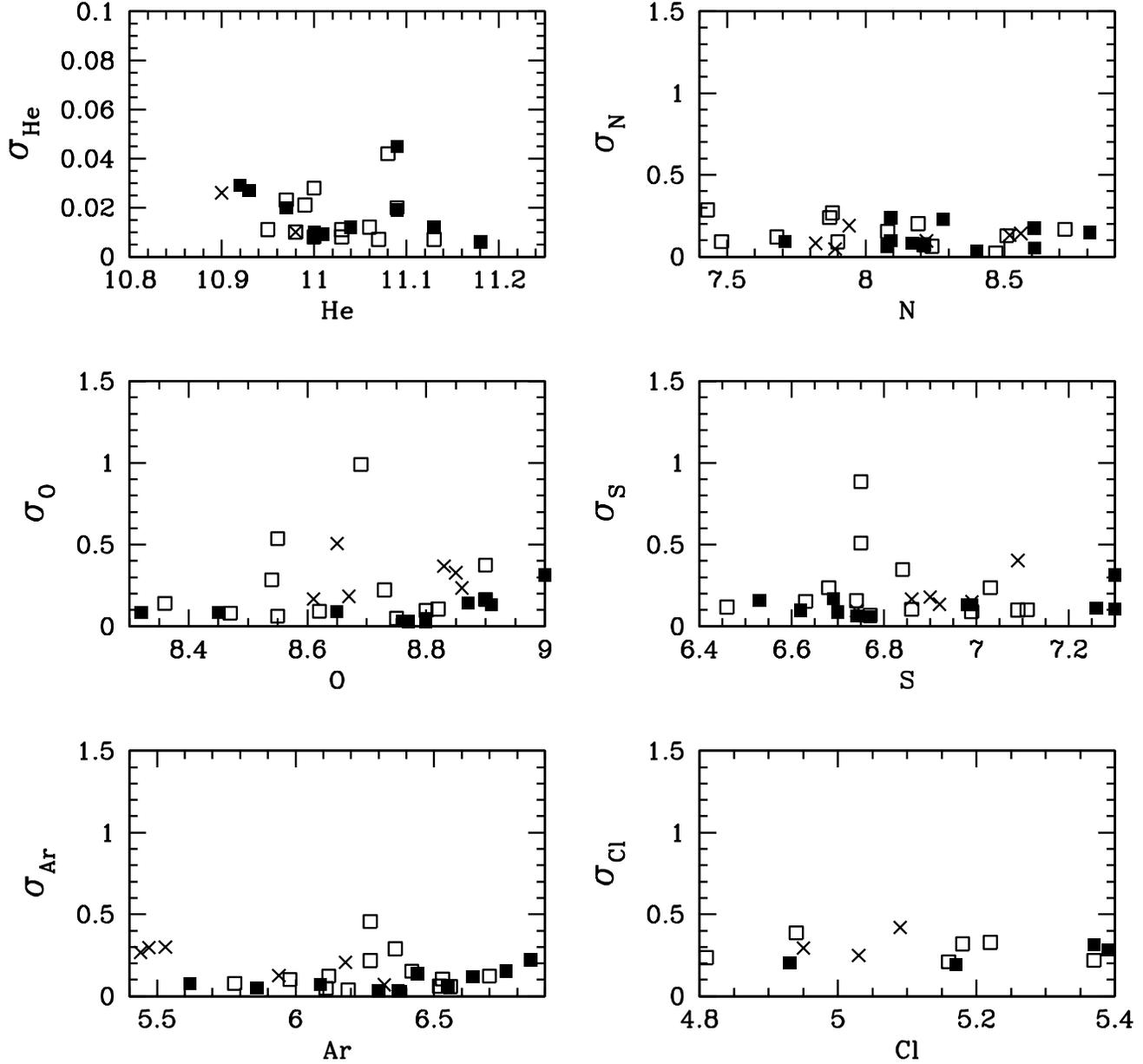}}
\caption{Estimated uncertainties on the derived abundances for the individual 
          PN. High excitation class PN (EC=6-7) are represented by filled squares,
medium excitation class PN (EC=4-5) by open squares, and low excitation class 
PN (EC$\leq 3$) by crosses. \label{figsigma}}
\end{figure*}

 Fig. \ref{figsigma} shows the resulting $1\sigma$ uncertainties on
the abundances as a function of the abundances themselves, for various
excitation classes.\\ 

\subsubsection{Uncertainties on the oxygen abundances}

Taking away the 3 PN with very high uncertainties, 
   a   systematic tendency  to an increase of the uncertainty on the
oxygen abundance for high oxygen abundances can be seen: if the uncertainty
is clearly lower than 0.2 dex for low oxygen abundances (O $\simeq$ 8.4--8.6),
it rises to much higher levels, as high as 0.4 dex for high oxygen
abundances (O $\geq$ 8.8). It is due to the fact that high oxygen abundance PN 
have low temperatures, and thus intrinsically weak [OIII]~436.3  lines, which 
are thus highly affected by the noise. This effect is clearly 
enhanced in comparison with Disk PN, due to  the high redenning of Bulge
PN.\\ 
  
It can be seen as well that the uncertainties go down as a general
tendency with increasing excitation class.\\

The 3 high uncertainty PN, as far as the oxygen abundance is concerned, are
 PN~G~001.4~+~05.3  (H1~-~15),  PN~G~358.2~+~04.2  (M3~-~8), and 
 PN~G~359.1~-~02.3  (M3~-~16).
These 3 Planetary Nebulae could not have the important diagnostic line
[OIII] 436.3~nm measured.

{\em PN G 001.4+05.3} 
 
This Planetary Nebula has as well the [NII] 575.5~nm auroral line at a very
low level. For this Planetary Nebula, it is the only temperature diagnostic
line that is present. Thus, the uncertainty on the electron temperature
in this PN is very high, as can be seen in Table \ref{tableplasma}. 

\vspace{1cm}

{\em PN G 358.2+04.2}

This 
Planetary Nebula has a high interstellar reddening, and, which is worse,
it is badly determined (uncertainty of 0.3 dex) (see Table \ref{tableplasma}).
This uncertainty thus transmits itself  to the intrinsic individual line
intensities.

{\em PN G 359.1-02.3}

This Planetary Nebula is of low electronic temperature and density. The uncertainties on
 the derived parameters lie therefore on a higher level.\\

\subsubsection{Uncertainties on the nitrogen abundances}

No systematical tendency with the abundance can be seen. As well, no
systematical tendency with the excitation class can be seen. The envelope
of the uncertainty level remains fairly constant at $\simeq 0.2$~dex.\\

\subsubsection{Uncertainties on the helium abundances}

Uncertainties on helium abundances show a clear tendency to decrease with 
increasing abundance. The emission mechanism for helium lines is mainly recombination. 
 Although collisional excitation and reabsorbtion 
can have  a secondary influence for the He~I
lines, the helium abundance is  basically a function
of the ratio of the helium lines to the hydrogen lines. The S/N ratio
 increasing with the  intensities of helium lines
(relative to ${\rm H\beta}$),  the uncertainty on the helium abundance
should decrease with increasing abundance, which is exactly what is observed.\\

The mean uncertainties on the helium abundances from our sample are thus
0.03--0.02 dex, following the helium abundance.\\

Two PN have higher uncertainties on their helium abundances, of the order
of 0.04 dex:  PN~G~357.4~-~03.2  (M2~-~18), and  PN~ G~ 358.2~+~04.2
  (M3~-~8). 
                 
The high uncertainty on the He abundance of  PN~G~358.2~+~04.2 is due to the
high uncertainty of its reddening, as is the case for the uncertainty
on its oxygen abundance.

On the other hand, the uncertainty on the helium abundance of   PN~G~357.4~-~03.2
 is certainly 
over-evaluated: its high value is due to the fact that it is one
 of the only PN in the sample with measured He I 402.6 and 728.1~nm lines. Due to their 
faint intensities, these lines  
 are  measured in low S/N conditions. Therefore, they enhance the  level
of the uncertainty. We derived as well the He abundance for this PN without these two 
lines, and it did not change significantly.

\subsubsection{Uncertainties on the sulphur, argon and chlorine abundances}

The uncertainties on the sulphur and argon abundances generally follow, 
although not as clearly, the tendencies of the 
uncertainties on the oxygen abundances, with an envelope of the uncertainties 
of 0.2--0.3 dex. This is due to the chain of ICFs leading to the elemental
abundances of sulphur and argon.\\ 

As well, the few Planetary Nebulae that are well above these uncertainties 
are the same as in the case of oxygen, for the same reasons.\\

It should be added that the uncertainties on
sulphur are clearly underevaluated, because of the inadequacy of the
sulphur ICF, as quoted in Cuisinier et al. (1996). The uncertainty on
the argon abundance should as well be underevaluated, to a lesser extent:
The ICF on argon shows a systematic effect of $\simeq$ 0.1dex (K\"oppen et al.
1991).\\

Finally, the chlorine abundances lie at higher uncertainty levels than the 
other ones, at 0.3--0.4 dex, with no systematic tendency, neither with 
abundance, nor with excitation class. This is due to the faintness of 
the chlorine lines --- which are measured at much lower S/N ratios than the 
other ones.

Furthermore, the levels of the  lines are so
low that systematic effects should as well be present,
 over--evaluating their intensities, as quoted in Rola \& Stasinska (1994).

\subsection{Comparison with other data}

There are five sources of Planetary Nebulae abundances determinations
 that present common data with our sample: 
 Aller \& Keyes  (1987), Costa et al. (1996), K\"oppen et al.
(1991), Ratag et al. (1997), and Webster (1988).
The comparison with these four samples will be discussed below.

K\"oppen et al. (1991) spectra are of survey nature, and generally therefore of  low quality.
They will thus not be further discussed here.

Ratag et al. (1997) joined observations of them, made on various sites, with already existing
abundances, in order to build a sample of Galactic Bulge Planetary Nebulae.
However, their abundances compare very badly with ours. The comparison is made dificult by
the lack of common objects with measured important diagnostic lines, like [OIII]~436.3nm. 

However, the reasonable comparison with other good samples from the literature (Webster 
1988, Aller \& Keyes 1987, Costa et al. 1996) 
gave us some  confidence in our parameters  and abundance 
determinations.\\

\begin{figure}
\resizebox{\hsize}{!}{\includegraphics{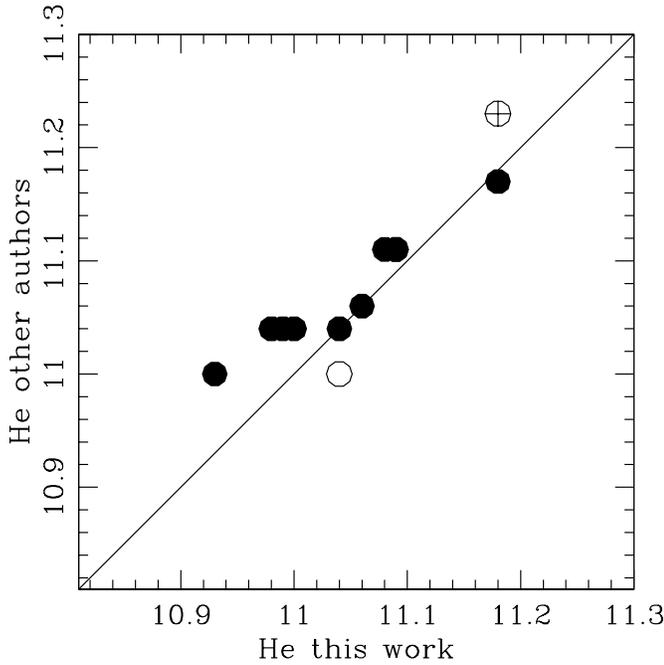}}
\caption{Comparison of our determinations of the helium abundances with  
determinations from other
authors. The helium abundances are represented on a logarithmic scale, where 12 is the 
hydrogen abundance.
 Filled circles stand for data from Webster (1988), open circles for data from Aller
\& Keyes 1987 and the cross for the common PN with Costa et al. (1996)   \label{figcomphe}}
\end{figure}

\begin{figure}
\resizebox{\hsize}{!}{\includegraphics{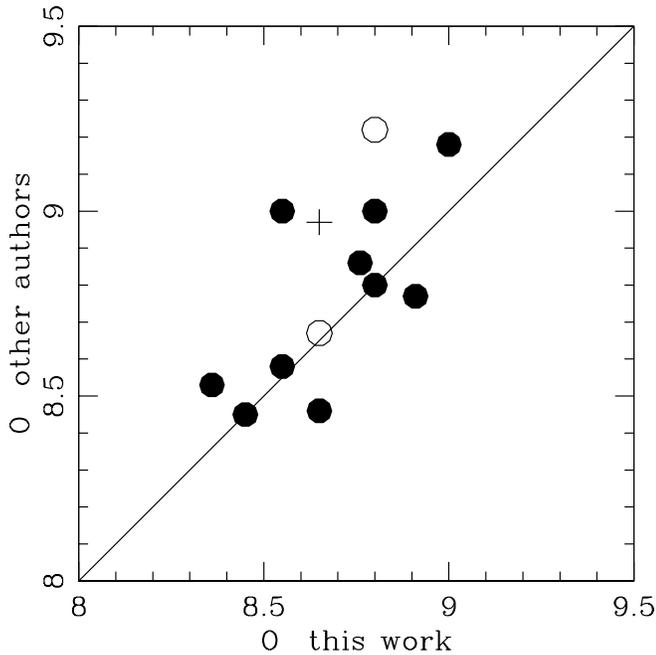}}
\caption{Comparison of our determinations of the oxygen abundances 
with  determinations from other 
authors.   
 The symbols are the same than in Fig. \ref{figcomphe} \label{figcompo}}
\end{figure}

\begin{figure}
\resizebox{\hsize}{!}{\includegraphics{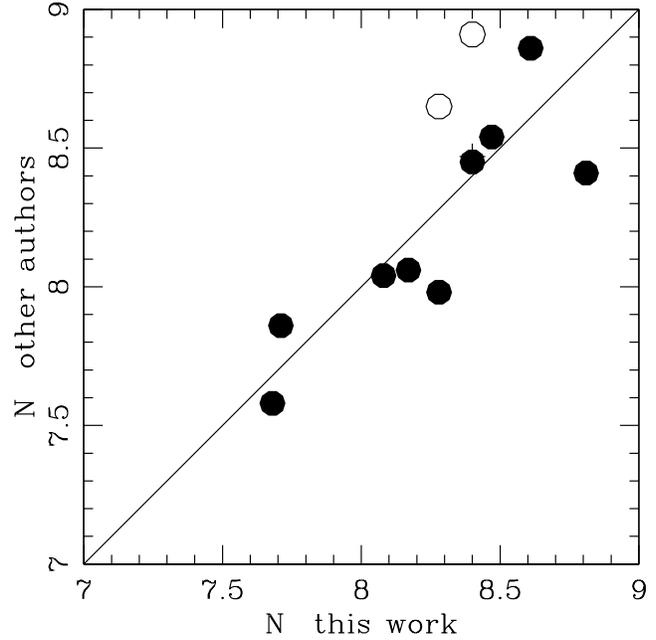}}
\caption{Comparison of our determinations of the nitrogen abundances 
with  determinations from other
authors.  
 The symbols are the same than in Fig. \ref{figcomphe}
   \label{figcompn}}
\end{figure}

We have 11 objects in common with Webster (1988),  2 with Aller \& Keyes (1987) 
(PN~G~002.7~-~04.8 =  M~1~-~42 and  PN~G~356.2~-~04.4 =  Cn~2~-~1)
 and 1 with Costa et al. (1996)
(Cn~2~-~1). Fig. \ref{figcomphe} shows the comparison
of their data with ours for the helium abundance, Fig. \ref{figcompo} 
for the oxygen abundance and \ref{figcompn} for the nitrogen abundance.\\

\subsubsection{Helium abundances}

A slight systematic tendency exists for helium, our abundances being lower by $\simeq 0.01$
dex than theirs, especially Webster's ones --- but this systematic tendency is well within our estimation of the uncertainty on the abundances  of our data. The dispersion around this mean tendency is of 0.01 dex as well.\\

\subsubsection{Oxygen abundances}

Also for oxygen, there is a slight systematic tendency, our abundances 
being systematically lower. The mean difference is of $\simeq 0.1$ dex, here
as well, well within our estimation of the uncertainty on the oxygen abundances.

It should be added that this systematic tendency clearly increases with the
 oxygen abundance, being much stronger for oxygen rich PN. Although only 
3-4 objects are  concerned, the difference rises up to 0.2-0.3 dex
for O/H$>9$ ({\em literature} abundances). 

If those objects are removed, the systematic tendency nearly
disappears. This could indicate a systematic overevaluation of the oxygen
abundance for oxygen rich PN from other studies. However, we cannot be
conclusive about this point, because of the small number of objects involved,
and because our evaluation of the random uncertainty
for such objects is of the same order, or even higher.

For two of these object (PN~G~000.7~+~03.2 =  M~2~-~21 and 
 PN~G~358.2~+~04.2 =  M~3~-~8),
  the  
usual diagnosis line [OIII]~436.3  could not be measured.
  Furthermore the [NII]~575.5 line, on which the electron temperature evaluation of these PN rely,  
was measured in such low S/N conditions that the error bars on 
the oxygen abundances are huge (respectivaly 0.3 and 0.5 dex!).  Thus, the 
increase of the systematic tendency  with oxygen abundance might not be 
so high.

 Cn~2~-~1 ( =  PN~G~356.2~-~04.4)  on the other hand has been observed  
in relatively good conditions
by Aller \& Keyes (1987), with measurements of [OIII]~436.3 and [NII]~575.5
at a reasonable S/N ratio. These lines are also present in our
observation  of this Planetary Nebula,  at a reasonable S/N and
the oxygen abundance derivations are very different (0.8 dex!).
 We attribute the difference to the contamination of 
[OIII] 436.3nm line by Hg 435.9 from the city lights 
in Aller \& Keyes observations. 
This contamination affects   their deconvolution
 and measurement of the [OIII] 436.3 line, and hence  their evaluation of 
the [OIII] electron temperature.   

The observation of M~1~-~42 by Costa et al. (1996) also presents   a higher 
oxygen abundance. We attribute this to their underestimation  of the reddening,
leading to an overevaluation of the intrinsic [NII] 658.4, 654.8 nm lines 
intensities. Thus, their [NII] electron temperature is underevaluated. Because  they did not 
observe [OIII] 436.3 nm, they adopted the [NII] temperature for the [OIII] zone, leading
to an overevaluation of the ${\rm O^{++}}$ abundance, and hence of the oxygen abundance.
They also find  a  ${\rm O^{++}/O^+}$ ratio which very low, 
leading through the ICF to a correct nitrogen 
abundance.

Thus, taking away these objects, the dispersion of the oxygen abundance 
is of the order of 0.1--0.2 dex,
 well within our estimation of the random errors.\\

\subsubsection{Nitrogen abundances}

Looking at the data from the literature as a whole, our abundances compare within 0.2 dex  with the abundances deduced by these authors,
 well within the generally 
quoted accuracy of PN abundances determination in the literature.  
No systematic tendency can be seen.

However, the two PN observed in common with Aller \& Keyes 1987,  Cn~2~-~1
(= PN~G~356.2~-~04.4) and  M~1~-~42 ( =  PN~G~000.7~-~03.2), 
present systematically 
lower abundances (by $\simeq$ 0.4 dex).
    
The nitrogen abundance is lower in Cn~2~-~1 because of the uncertain deconvolution of [OIII] 436.3 nm, leading to an incorrect ${\rm O^{++}}$ abundance, that reflects itself in the nitrogen abundance through the ICF. 

The case of M~1~-~42 is a bit more tricky, 
because the nitrogen abundance is very different, 
whereas the oxygen abundance is very similar (in Aller \& Keyes 1987). 
This can be explained by the fact that 
the [NII]~575.5~nm is {\em overestimated} by a factor of 10, leading to a very
high [NII] temperature (in their observations). 
This combines with the problems in the deconvolution
of [OIII]~436.3~nm. The measurement of this line is even qualified as uncertain 
by the authors, and therefore may  well be underestimated, leading to an 
underestimation of the [OIII] temperature, and to an overestimation of the 
[OIII] zone abundances. The combination of these two effects gives a correct
oxygen abundance, but overevaluates the nitrogen abundance.
 
\section{Abundances distributions and discussion}

\subsection{Oxygen, sulphur and argon abundances}

Oxygen, sulphur and argon are not synthetised in the progenitors of
the Planetary Nebulae. Thus, these elements reflect the composition of the 
interstellar medium at the time when the progenitor star was born 
--- and allow us to 
trace its chemical evolution.\\

In order to see wether the Bulge was chemically more or less evolved than 
the Disk, from the point of view of PN,
we compared the abundances of these elements in our Bulge  sample 
with the abundances of a Disk PN sample,  taken from 
Maciel \& K\"oppen  (1994). 
We took this sample as a ``best compromise'', because the Disk does clearly
not constitue  one unique population, but rather   several ones. The 
variety of populations of the Disk reflects itself in the radial abundances
gradients (Maciel \& K\"oppen 1994, K\"oppen et al. 1991, Fa\'undez-Abans \&
Maciel 1986). 
The effect of these gradients is to smooth and spread the abundance
distributions over the Disk, but they do not change the form of the 
Disk abundance distributions as a whole. 
We resolved to adopt the abundances published in Maciel \& K\"oppen 1994,
because they represent the most comprehensive sample of PN abundances to this
date. Through this choice, we hope to minimize systematic effects due to the 
choice of PN at a particular place in the Galaxy rather than at another one.\\

\begin{figure}
\resizebox{\hsize}{!}{\includegraphics{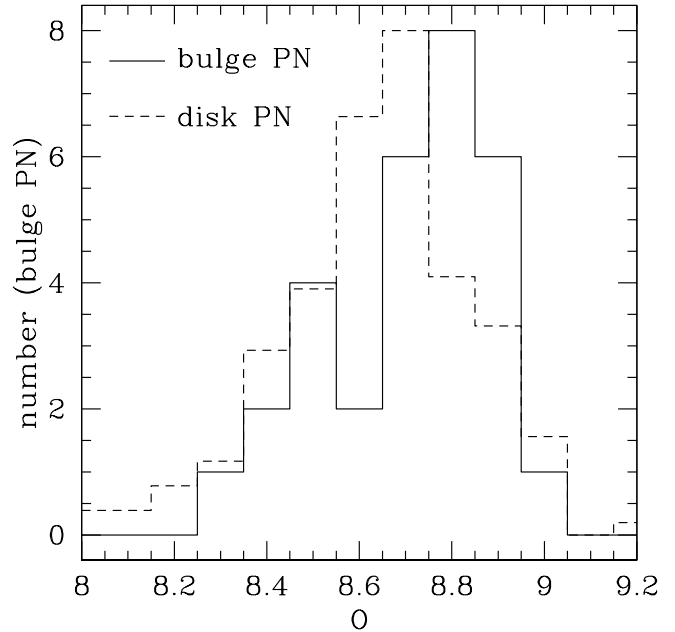}}
\caption{Oxygen abundances for  PN in the Bulge and in the Disk. The Bulge 
sample contains 30 PN, the Disk sample 198 PN. The vertical scale shows the 
number of objects in the Bulge sample.}
 \label{fighisto}
\end{figure}

We found that the oxygen abundances  in the Bulge PN are comparable to the 
abundances in oxygen in the Disk PN, the high abundance PN in the Bulge
(O/H$\simeq$8.8--9.0) being slightly more numerous, as 
can be seen in Fig.  \ref{fighisto}.
The mean abundance we encountered in the Bulge (${\rm <0/H>=8.71}$) is also similar to 
the mean abundances that can be encountered in the youngest populations of the Disk, either
in type I PN in the Disk (Costa et al. 1996), or in HII regions or B stars (Shaver et al. 1983,
Smartt \& Rolleston 1997, Maciel \& Quireza 1999).\\

From the stellar point of view, it seems that the very high metallicities that were 
encountered in the Bulge from low resolution data in earlier studies (up to 
${\rm [Fe/H]\simeq +1 dex}$)(Whitford \& Rich 1983, Rich 1988, Geisler \& Friel 1992)
were really extrapolated too far. High resolution abundance studies have shown
that the highest iron abundances that can be encountered, either in the Disk or in the
Bulge giants, are of the order of ${\rm [Fe/H] \simeq +0.6 dex}$ (Mc~William \& Rich 1994,
Castro et al. 1995,1996,1997). Mc~William \& Rich (1994) rescaled the low resolution study
of Rich (1988), and encountered an iron abundance distribution that is similar to that of 
Mc William (1990) Disk giants sample, with a mean abundance of ${\rm <[Fe/H]>=-0.23 dex}$. 
Within errors, this is similar to the mean iron abundance that can be encountered from G--dwarfs in the 
solar neighbourhood (Pagel 1989, Rocha-Pinto \& Maciel 1996). A closer inspection of the data
shows that if the mean abundances are similar, the high abundance stars (${\rm [Fe/H] \geq 0 dex}$) 
are slightly more frequent in the Bulge than in the solar neighbourhood G-dwarfs. 
 The solar neighbourhood G-dwarfs can be considered to be more representative of the
Disk iron abundances than Mc William giants, because Mc William giants were selected on luminosity 
criteria, and are therefore biased in favour of massive stars, that should be slightly more 
metal-rich, as stated in Mc~William \& Rich 1994.

Recently, Sadler et al. (1996) made a new low resolution spectroscopical survey of a much larger 
sample than Rich (1988), and they encountered a slightly higher value of the mean iron abundance
(${\rm <[Fe/H]>=-0.11 dex}$). However, they have a significant fraction of very high metallicity
stars (${\rm [Fe/H] \geq 0.6 dex}$), that pushes their mean iron abundance to high values. We thus prefer
the Rich (1988) sample, rescaled to Mc~William \& Rich (1994) values, which we find to be  
in better agreement with the iron upper abundance limit from high resolution studies.

Thus, the oxygen abundances in PN behave like the iron abundances in stars, when the Bulge and the Disk
populations are compared: the distributions are similar, high abundance objects being slightly
over-numerous  in the Bulge.\\

This is somehow unexpected from a theoretical point of view: the rapid star formation history of
the Bulge, as would be expected for an old population, should enhance the nucleosynthesis
products of supernovae of type II over thos of supernovae of type Ia, as the $\alpha$~elements
(Matteucci \& Brocatto 1990). Thus, oxygen should be enhanced over iron in the Bulge, 
contrarely to what seems to appear from the comparison of them stellar and the PN abundances
in the Bulge and in the Disk. The same effect appears in the  Mc William \& Rich (1994)
high resolution observations of Bulge giants. they found solar values for their
[O/Fe] ratios. More generally, the $\alpha$~element pattern that they encountered 
in the Bulge is rather puzzling: some of their $\alpha$~elements are indeed enhanced,
(like Mg and Ti), as would be expected from a rapid evolution, and others are not
(O, Ca and Si). Idiart et al. (1996) also encountered  an enhanced magnesium
abundance in their integrated spectrum in Baade's window.\\

Aluminium is also enhanced in Mc William \& Rich Bulge giants, thus the possibility
of oxygen depleting processes should not be discarded. The results of such depleting 
processes have actually been observed in some globular cluster giants (Kraft et al. 1995).
However, as stated in Richer et al. (1998), such processes are very unlikely to have
occurred in the in the Bulge PN progenitors, because they would also have dramatically 
enhanced the nitrogen abundances, whereas as we will see in the next section, our Bulge PN 
have {\em low} N/O ratios, and never reach the values observed in oxygen depleted giants.
(${\rm [N/O]\geq 1}$, e.g. ${\rm N/O \geq 0.1}$, adopting Anders \& Grevesse (1989) values 
for the solar abundances, according to Kraft et al. (1995) or Denissenkov et al. (1998)).
Thus, we can consider that the oxygen abundances that we encountered in the Bulge PN have 
not been polluted by their progenitor nucleosynthesis, and really represent pristine 
material.

We also analysed  the abundances in  sulphur and argon.  As in the Disk PN
(Fa\'undez--Abans \& Maciel, 1986, Cuisinier et al. 1996), we found them 
well correlated with the oxygen abundances.  However, these abundances,
 suffer from higher uncertainties than the oxygen abundances  
due to the uncertainties on their ICFs.
 Within these uncertainies, they do not contradict the
results from the oxygen abundances.\\

\subsection{Helium and nitrogen abundances}

Helium and nitrogen, contrarely to oxygen, sulphur, and argon, are synthetized 
in the progenitors of PN.
 Thus, these elements do not allow us to trace  the chemical
composition of the interstellar medium when the progenitor star was born,
and hence they do not allow us to trace the chemical evolution of the 
interstellar medium.

 On the other hand, and although it is not yet well 
quantified theoretically,  the enrichment in these elements 
should be connected with the mass of the progenitors, and thus, with their
ages (Cazetta \& Maciel 1999, Stasinska \& Tylenda 1990,  Kaler 1983,1985,
Peimbert \& Torres--Peimbert 1983, Aller 1987 et al.).\\

The N/O ratio abundance ratio distribution should therefore reflect a mixture of  
the chemical evolution of nitrogen and oxygen before progenitor stars were 
formed, (2) the progenitor own nucleosynthesis.\\

It is noteworthy  that the highest metallicity (e.g. oxygen 
abundance) galaxies do not produce N/O ratios higher than ${\rm N/O \simeq -0.25}$
(Vila--Costas \& Edmunds 1993). Within errors, this is also the upper
limit of the N/O distribution of our sample (Fig. 6).

However, the Disk PN N/O ratio distribution extends to much higher
values --- due to the  nucleosynthesis of massive progenitor stars
(The Kolmogorov--Smirnoff probability that the N/O distributions are
different is 74\%). Thus, the most recent population of PN in the
Disk is clearly absent from the Bulge. This is what should be
expected from stellar observations, showing that the turnoff lies
around ${\rm V-I \simeq 0.6}$ (Ortolani 1998), and
is therefore an old population.\\

The helium abundances although not as clear, show a similar
effect, as can be seen in Figure 7: The most abundant objects 
that are present in in the Disk are absent in the Bulge.
The distribution  encountered in the Bulge compares with the
Disk PN, if the high helium abundance type I PN are excluded.

We see that the mean value that we encounter, 
${\rm <He/H>=0.108}$ (on a linear scale, considering only
our best data), is lower than the pristine helium abundance
predicted by chemical evolution models for the Bulge
(${\rm He/H \simeq 0.12}$)(Catelan \& de Freitas Pacheco 1996).
Helium abundances can only be obtained from indirect methods
for stars in the Bulge, like differential counts in color 
magnitude diagrams. Although they can therefore considered 
 to be not as trustworthy as the nebular abundances,
good quality data seem to favour low helium abundance values
(He/H=0.098)(Minniti 1995), although higher values 
have been reported from lower quality data (Renzini 1994).

It seems that here as well, helium does not follow the 
overproduction predicted by chemical evolution  models for a
prompt enrichment.

\begin{figure}
\resizebox{\hsize}{!}{\includegraphics{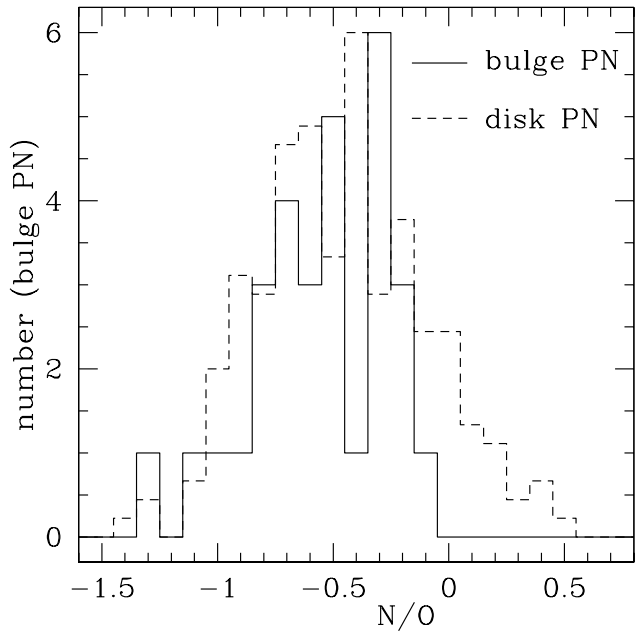}}
\caption{N/O ratio for PN in the Bulge  and in the Disk. The Bulge sample contains 30 PN, the Disk sample 198 PN. 
  \label{fighistno}  }
\end{figure}

\begin{figure}
\resizebox{\hsize}{!}{\includegraphics{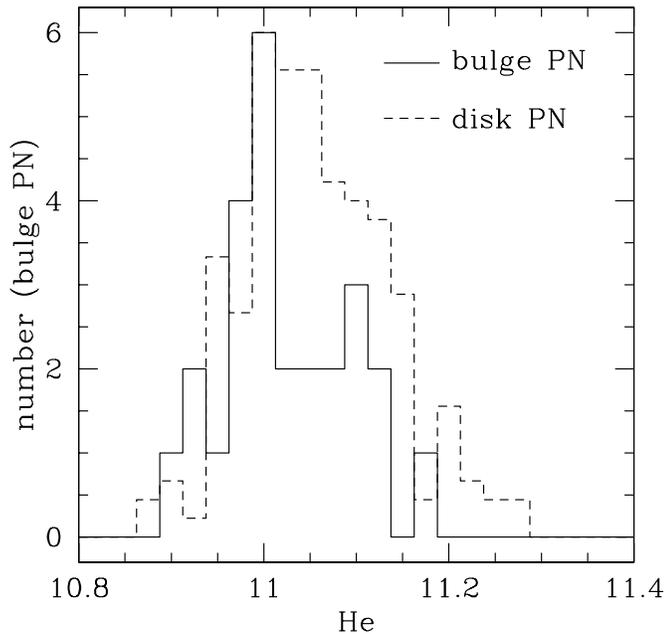}}
\caption{He abundances for PN in the Bulge and in the Disk.  The Disk sample contains 198 PN, the Bulge sample contains 17 PN (only PN with good helium abundances determinations, with EC$\geq$4 have been considered).
 \label{fighisthe}}
\end{figure} 

\section{Conclusions}

We observed a sample of 30 Planetary Nebulae in the Galactic Bulge with high quality spectroscopy, 
for which we derived plasma parameters, and abundances  of
 O, S, Ar,Cl, N, and He.\\

We evaluated random uncertainties with appropriate simulations, which were 
for these highly reddened Planetary Nebulae of the order of
0.02 dex for helium, of 0.2 dex for oxygen, sulphur, and argon, and of 
0.4 dex for chlorine.

We compared the abundances we obtained with abundances derived by other 
authors, and our abundances fitted well within our estimations of the 
uncertainties.\\

We compared the abundances in O, S, Ar of the Planetary Nebulae  from our Galactic Bulge sample 
 with the  abundances in Planetary Nebulae  from the Galactic Disk. We found them to be comparable,  
  Planetary Nebulae  with high abundances in these elements    being
slightly more numerous --- which is compatible with the findings from the
 stars.

The abundances in N/O and He showed a lack of high N/O and high He Planetary Nebulae 
in the Galactic Bulge, suggesting that the Bulge Planetary Nebulae  have old progenitor stars, 
and here as well are similar to the stellar population. 

Thus, the main results that appear from our observations are that:
(1) The abundances in O, S, and Ar do not contradict the stellar observations 
in the Bulge
(2)
From the N/O ratio derivations, it seems that the Bulge Planetary Nebulae  are an old population, like the stars.\\

If the abundances that can be derived from stars and Planetary Nebulae in the Galactic Bulge are not contradictory,
the pattern that is found can still be considered as very puzzling: if there is
no doubt that the most recent population that is encountered in the Galactic Disk does not 
appear either in the Planetary Nebulae or in the stars, some elements favour
a prompt chemical enrichment (Mg and Ti), and other favour a much slower chemical evolution
history (He, O, Si, S, Ar and Ca).

\acknowledgements 

F.C. thanks FAPESP for financial support, through the grant 95/04766--8, and 
the 
Instituto Astron\^omico e Geofisico da USP, 
for hospitality during his stay as a post--doctoral
fellow. W.J.M. thanks FAPESP and CNPq for partial financial support.

We would like to thank as well the referee, J.A. de Freitas Pacheco, 
for helpful suggestions, and G. Stasinska, for fruitful discussions.


\begin{thebibliography}{}
\bibitem{}   Acker A., K\"oppen J., Stenholm B., Raytchev B., 1991,
             A\&AS, 89, 237
\bibitem{}   Acker A., Ochsenbein F., Stenholm B., et al., 1992,
               ``The Strasbourg--ESO catalogue of galactic
                     Planetary Nebulae'', ESO publ.
\bibitem{}   Aller L., Keyes C., Maran S., 1987, ApJ, 320, 159
\bibitem{} Aller L., Keyes C., 1987, ApJS, 65, 405
\bibitem{} Anders E., Grevesse N., 1989, Geochimica \& Cosmochimica Acta,
           53, 197
\bibitem{} Baldwin J.A., Stone R.P.S., 1984, MNRAS, 206, 241
\bibitem{} Castro S., Barbuy B., Bica E. et al., 1995, A\&AS, 111, 17
\bibitem{} Castro S., Rich R.M., Mc William et al., 1996, AJ, 116, 2439
\bibitem{} Castro S., Rich R.M., Grenon M. et al., 1997, AJ, 114, 376
\bibitem{} Catelan M., de Freitas Pacheco J.A., de Fran\c ca Jr. J.A., 1996, A\&A, 313, 924
\bibitem{} Cazetta, J.O., Maciel W.J., 1998, in preparation
\bibitem{} Colina L., Bohlin R., 1994, AJ, 108, 1931
\bibitem{} Costa R.,  de Freitas Pacheco J.A., de Fran\c ca Jr., J.A., 1996,
             A\&A, 313, 924 
\bibitem{} Cuisinier F., Acker A., K\"oppen, 1992, in ``The Feeedback of
              chemical evolution on the stellar content of Galaxies'', p. 99
\bibitem{} Cuisinier F., Acker A., K\"oppen J, 1996, A\&A, 307, 215
\bibitem{} Denissenkov P.A., Da Costa G.S., Norris G.E., Weiss A., 
           1998, A\&A, 333, 928
\bibitem{} Fa\'undez-Abans M., Maciel W., 1986, 158, 228
\bibitem{} Hamuy M., Walker A., Suntzeff N., et al., 1992, PASP, 104, 533
\bibitem{} Hamuy M., Suntzeff N., Heathcote S., et al., 1994, PASP, 106, 566
\bibitem{} Idiart T., de Freitas Pacheco J.A.,  Costa R.D.D., 1996, AJ, 111, 1169
\bibitem{} Kaler J., 1983, ApJ, 271, 188
\bibitem{} Kaler J., 1985, Ann. Rev. A\&A, 23, 89
\bibitem{} K\"oppen J., Acker A., Stenholm B., 1991, A\&A, 248, 197
\bibitem{} Kraft R.P., Sneden C., Langer G.E. et al., 1995, AJ, 109, 2586
\bibitem{} Maciel W., K\"oppen J., 1994, A\&A, 282, 436
\bibitem{} Maciel W., Chiappini C., 1994, Ap\&SS, 219, 231
\bibitem{} Maciel W., Quireza C., 1999, A\&A, 345, 629
\bibitem{} Matteucci F., Brocatto E., 1990, ApJ, 365, 539
\bibitem{} Minniti D., A\&A, 300, 109
\bibitem{} Mc William A., 1990, ApJS, 74, 1075
\bibitem{} Mc William A., Rich R.M.,  1994, ApJS, 91, 749
\bibitem{} Massey P., Strobel K., Barnes J., Anderson E., 1988, ApJ, 328, 31
\bibitem{} Oke J., 1990, AJ, 99, 1621
\bibitem{} Ortolani S.,  1998, in ``Connecting the Distant Universe with the Local
Fossil Record'', Spite \& Spite eds.
\bibitem{} Pagel B.E.J., 1989, in ``Evolutionary phenomena in galaxies'', Beckman, Pagel eds.,
              Cambridge University Press, p. 201
\bibitem{} Pagel B.E.J., 1997, ``Nucleosynthesis and Chemical Evolution of Galaxies'',
Beckman, Pagel Eds., Cambridge University Press
\bibitem{} Peimbert M., Peimbert S., 1983, IAU symp. 103, p. 233
\bibitem{} Ratag M., Pottasch S., Dennefeld M., Menzies J., 1992, A\&A, 255, 
  255
\bibitem{} Ratag.M., Pottasch S., Dennefeld M., Menzies J., 1997, A\&AS, 126, 
 297  
\bibitem{} Renzini A., 1994, A\&A, 285, L5
\bibitem{} Rich R.M., 1988, A.J., 95, 828
\bibitem{} Richer M.G., Mc Call M.L., Stasinska G., 1998, A\&A, 340, 67
\bibitem{} Rola C., Stasinska G., 1994, A\&A, 282, 199
\bibitem{} Rocha--Pinto, H., Maciel W., 1996, MNRAS, 279, 447
\bibitem{} Sadler E.M., Rich R.M., Terndrup D.M., 1996, AJ, 112, 171
\bibitem{} Shaver P.A. Mc Gee R.X., Newton L.M. et al., 1983, MNRAS, 204, 53
\bibitem{} Smartt S.J., Rolleston W.R.J., 1997, ApJ, 481, L47
\bibitem{} Stasinska G., Tylenda R., 1990, A\&A, 240, 467
\bibitem{} Stone R.P.S., Baldwin J.A., 1983, MNRAS, 204, 347
\bibitem{} Stone R.P.S., 1977, ApJ, 218, 767
\bibitem{} Stone R.P.S., 1996, ApJS, 107, 423
\bibitem{} Vila--Costas M.B., Edmunds M.G., 1993,  MNRAS, 265, 199
\bibitem{} Webster L., 1988, MNRAS, 230, 377 
\bibitem{} Whitford A.E., Rich R.M., 1983, ApJ, 274, 723
\end{thebibliography}
\end{document}